\documentstyle[twoside,fleqn,espcrc2]{article}
\newlength{\bredde}
\def\slash#1{\settowidth{\bredde}{$#1$}\ifmmode\,\raisebox{.15ex}{/}
\hspace*{-\bredde} #1\else$\,\raisebox{.15ex}{/}\hspace*{-\bredde} #1$\fi}

\newcommand{\beq}{\begin{equation}}
\newcommand{\eeq}{\end{equation}}
\def\beqn{\begin{eqnarray}}
\def\eeqn{\end{eqnarray}}

\newcommand{\AmS}{{\protect\the\textfont2
  A\kern-.1667em\lower.5ex\hbox{M}\kern-.125emS}}

\title{The QCD Dirac Operator Spectrum and Finite-Volume Scaling}

\author{P.H. Damgaard\address{The Niels Bohr Institute,
Blegdamsvej 17, DK-2100 Copenhagen, Denmark}%
        \thanks{Work supported by EU TMR Grant no. ERBFMRXCT97-0122.}}
       
\begin{document}

\begin{abstract}
Random matrix theory and chiral Lagrangians offer a convenient tool for the 
exact calculation of microscopic spectral correlators of the Dirac operator
in a well-defined finite-volume scaling regime.
\end{abstract}

\maketitle

\section{INTRODUCTION}

The succesful analytical computations of the so-called microscopic
QCD Dirac operator spectrum from random matrix theory \cite{SV,VZ1,DN1}
have raised an important question. Why can random matrix theory, which
seems to be very far from the quantum field theory of QCD, apparently
be used to compute spectral properties of the Dirac operator?
Within the last year this issue has been much clarified, with important 
consequences also for lattice gauge theory. The results obtained in random 
matrix theory are now seen to be exact predictions of QCD itself, 
in an appropriate finite-volume scaling regime. This is defined by \cite{LS}
$1/\Lambda_{QCD}\! <<\! L\! <<\! 1/m_{\pi}$, where $L\!=\!V^{1/4}$ is the linear
extent of the four-volume $V$. In this region one is probing 
finite-volume effects of QCD to the extreme. Only static modes of the lowest
hadronic excitations contribute to the euclidean partition function of
QCD, and if chiral symmetry is spontaneously broken according to
$SU(N_f)\times SU(N_f) \to SU(N_f)$, it reads \cite{LS}
\beq
{\cal Z}_{\nu} = \int_{U(N_{f})} dU (\det U)^{\nu} e^{V\Sigma 
ReTr[{\cal M}U^{\dagger}]} \label{ZLS}
\eeq
in a sector of fixed topological charge $\nu$. Here $\Sigma\!=\!
\langle\bar{\psi}\psi\rangle$ is the quark condensate, and $\cal{M}$
is the quark mass matrix. In this context it is most useful to view
${\cal Z}_{\nu}$ 
as the generating function of $\bar{\psi}\psi$ (because this is 
how one sees that it contains information about the Dirac operator spectrum).
It depends on the masses only in the particular combination
$\mu\!=\!mV\Sigma$. This makes ${\cal Z}_{\nu}$ a finite-size scaling 
function.

\subsection{The double-microscopic limit}

Eigenvalues $\lambda_n$ of the Dirac operator are determined by
$i\,\slash{D}\phi_n\!=\!\lambda_n\phi_n$. By comparing (\ref{ZLS}) with the 
original path integral (after integrating out the fermions), 
\beq
{\cal Z}_{\nu} = \prod_f m_f^{\nu} \int [dA]_{\nu} \prod_{f,n}
(\lambda_n^2+m_f^2)e^{-S[A]} ~,
\eeq
one infers immediately that also the non-zero eigenvalues $\lambda$
in this finite-volume limit can only enter in the combination 
$\zeta\!=\!\lambda V\Sigma$. It is thus a {\em double-microscopic}
limit $\lambda, m \to 0$ as $V\!\to\!\infty$ with $\mu$ and $\zeta$
fixed that will correspond to this finite-volume scaling regime
\cite{SV,JNZ,DN1}. Remarkably, in this limit random matrix theory 
can be used to give am entirely equivalent description of the partition
function (\ref{ZLS}). The relevant random matrix theory is given by \cite{LS}
\beq
\tilde{\cal Z}_{\nu} 
~=~ \int\! dW \prod_f{\det}\left(M + m_f\right)~
e^{-\frac{N}{2} tr~ V(M^2)} 
\label{ZRM}
\eeq
where
\beq
M ~=~ \left( \begin{array}{cc}
              0 & iW^{\dagger} \\
              iW & 0
              \end{array}
      \right) 
\eeq 
and $W$ is a rectangular complex matrix of size
$N\!\times\!(N\! +\! |\nu|)$.
This matrix ensemble, which is the one relevant for QCD, is known as the chUE. 
The choice of
ensemble, and the determinantal structure in (\ref{ZRM}), 
is fixed by the symmetry properties of the Dirac operator \cite{V}, 
while nothing a priori fixes the matrix potential $V(M^2)$. For simplicity, 
denote also the eigenvalues of $M$ by $\lambda$. In ref. \cite{SV} it
was shown for the particular case of a Gaussian random matrix potential
that in the double-microscopic limit the two partition functions agree
(up to an irrelevant mass-independent constant):
\beq
{\cal Z}_{\nu}(\mu_1,\ldots,\mu_{N_{f}}) ~=~ 
\tilde{\cal Z}_{\nu}(\mu_1,\ldots,\mu_{N_{f}}) ~. 
\label{ZZ}
\eeq 
The required correspondences between the two formulations are $\Sigma
\leftrightarrow \pi\rho(0)$ and $V \leftrightarrow 2N$, where 
$\rho(\lambda)$ is the density of the random matrix eigenvalues.
One particular consequence of a series of universality theorems (see
ref. \cite{ADMN} and the first of ref. \cite{DN1}) is that this equality
holds {\em universally}, i.e. independently of the choice of random matrix
model potential \cite{D1}. The only essential requirement is, as one
could expect, that $\rho(0)\! \neq\! 0$. This means that
masses and eigenvalues can be rescaled to common values.\footnote{
It is tempting to consider the case of $\rho(0)\!\to\! 0$ in
random matrix language as the limit of chiral symmetry restauration,
but there is presently no obvious classification of quantum field theories 
that match the new random matrix theory universality classes with 
$\rho(0)\!=\!0$ \cite{ADMN1}.}

The universal equivalence 
(\ref{ZZ}) is the root from which all exact relations
between random matrix theory and QCD in the finite-volume scaling regime
$1/\Lambda_{QCD}\! <<\! L\! <<\! 1/m_{\pi}$ follow. Because there are 
long-established techniques for computing spectral correlation functions
in random matrix theory, this equivalence is not just interesting
from a conceptual point of view, but is of great practical value.

\section{DIRAC SPECTRA AND PARTITION FUNCTIONS}

One particularly simple way of computing the microscopic spectral
correlators of the Dirac operator eigenvalues is by relations to
partition functions with additional quark species. In random matrix theory
based on the chUE ensemble there exists a convenient shortcut for the 
calculation of spectral correlation functions which is based on the kernel 
$K_N(z,z';m_1,\ldots,m_{N_{f}})$. From this kernel alone one can derive
all spectral correlation functions in the limit $N\to\infty$:
\beqn
&&\rho(\lambda_1,\ldots,\lambda_n;m_1,\ldots,m_{N_{f}}) \cr
&&~~~=~~ \det_{a,b} K(\lambda_a,\lambda_b;m_1,\ldots,m_{N_{f}}) ~.
\label{correl}
\eeqn
This relation is exact, and not restricted to the double-microscopic
limit. When comparing the explicit expression for the kernel with the 
eigenvalue description of the random matrix partition function (\ref{ZRM}) 
itself, one finds a very simple relation involving the ratio of two
partition functions, one corresponding to having two additional quark
species of imaginary mass! In the double-microscopic limit this relation
simplifies even further, and if we now make use of the equivalence
(\ref{ZZ}), we can express this kernel entirely in terms of finite-volume
partition functions in the scaling region $1/\Lambda_{QCD}\! <<\! L\! 
<<\! 1/m_{\pi}$ \cite{D2}:
\beqn
&&\!\!\!\!\!\!\!\!\!K_S(\zeta,\zeta';\mu_1,\ldots,\mu_{N_{f}}) \!=\! \prod_{f}
\sqrt{(\zeta^2\!+\!\mu_f^2)(\zeta'^2\!+\!\mu_f^2)}\cr
&&\times\sqrt{\zeta\zeta'}\frac{
{\cal Z}_{\nu}(\mu_1,\ldots,\mu_{N_{f}},i\zeta,i\zeta')}{
{\cal Z}_{\nu}(\mu_1,\ldots,\mu_{N_{f}})} ~,\label{mf}
\eeqn
up to an overall normalization constant that can be fixed by matching
with the spectral density at the origin. Eq. (\ref{mf}) is a master formula
from which all microscopic spectral properties of the QCD Dirac operator
can be derived. In particular, the microscopic spectral density itself
is obtained by taking the coincident limit $\zeta\!=\!\zeta'$:
\beqn
&&\rho_S(\zeta;\mu_1,\ldots,\mu_{N_{f}}) \!=\!
|\zeta| \prod_{f}(\zeta^2+\mu_f^2)\cr &&\times\frac{
{\cal Z}_{\nu}(\mu_1,\ldots,\mu_{N_{f}},i\zeta,i\zeta)}{
{\cal Z}_{\nu}(\mu_1,\ldots,\mu_{N_{f}})} ~,\label{spec}
\eeqn
again up to an overall constant, which is easily fixed \cite{D2}. Although
it involves a theory with two more quarks, one can
also view it as the expectation value of a complicated operator
in the original theory with just $N_f$ quarks.

It turns out that also the distribution $P_S(\zeta;\mu_1,\ldots,\mu_{N_{f}})$ 
of just the smallest 
Dirac operator eigenvalue can be expressed in terms of the finite-volume
partition function. In the sector of topological charge $\nu\!=\!0$ it reads
\cite{NDW}
\beqn
&&P_S(\zeta;\mu_1,\ldots,\mu_{N_{f}}) = \cr
&&\!\!\!\!\!\!\!-\frac{\partial}{\partial\zeta}\left[e^{-\zeta^2/4}
\frac{{\cal Z}_0(\sqrt{\mu_1^2\!+\!\zeta^2},\ldots,
\sqrt{\mu_{N_{f}}^2\!+\!\zeta^2})}{{\cal Z}_0(\mu_1,\ldots,\mu_{N_{f}})} 
\right]
\label{small}
\eeqn
and similar expressions can be derived for any $\nu$.

The expressions (\ref{spec}) and (\ref{small}) for the microscopic spectral
density and smallest eigenvalue distributions are the most convenient
quantities to measure in lattice gauge theory. Their analytical expressions,
using the explicit form of the finite-volume QCD partition function \cite{JSV},
agree precisely with results obtained earlier based directly on random matrix
theory computations \cite{VZ1,DN1}. Since the analytical results are known
for arbitrary masses (keeping the condition  $1/\Lambda_{QCD}\! <<\! L\! 
<<\! 1/m_{\pi}$), they can in principle be measured by Monte Carlo
techniques without difficulty even with dynamical fermions. Existing
Monte Carlo meaurements \cite{B} for QCD in 4 dimensions are based on
staggered fermions and an $SU(2)$ gauge group, for which predictions
actually are slightly different. Very recently the expression (\ref{spec})
has been derived directly from the chiral Lagrangian framework, without
relying on the relation to random matrix theory at all \cite{OTV}.

\subsection{QCD in 3 dimensions}

One can also entertain the idea of spontaneous breaking of ``chiral'' 
symmetries in 3-dimensional QCD. Because fermions are described by
two-spinors, and $\gamma$-matrices thus can be taken as the 
Pauli matrices (and {\em minus} these matrices, a different representation
of the Clifford algebra here), there is no room for an analogue
of $\gamma_5$. However, by pairing up quark fields one can reformulate
the theory in terms of four-spinors. Two hidden chiral symmetries (based
on $\gamma_4$ and $\gamma_5$ rotations) then become manifest. For an
even number of flavors it is conceivable that flavor symmetry, in the
guise of chiral symmetry,
can break spontaneously according to $U(N_f)\!\to\!U(N_f/2)\!\times\!
U(N_f/2)$, a case that has been analyzed in the random matrix framework
in \cite{VZ2}. It recently been studied on the lattice
in a quenched simulation, and indeed a condensate was observed at zero 
temperature, disappearing at the deconfinement phase
transition \cite{DHKM}. In the broken phase one can study the microscopic
Dirac operator spectrum, which turns out to be somewhat different from
the 4-dimensional counterpart.

The case of quenched 3-dimensional QCD is the only known case for which the 
prediction for the microscopic spectral density is exactly {\em flat}.
New Monte Carlo results on lattices of sizes ranging from $6^3$ to $14^3$,
show fairly good agreement with this prediction \cite{DHKM}. 
Getting a flat microscopic distribution is in fact quite non-trivial, 
since generically the spectral density reflects the underlying peaks 
corresponding to the averaging over individual peaks of the smallest 
eigenvalues. It will of course be more interesting to see the result
of an analogous simulation with dynamical fermions, for which
the analytical prediction is a highly non-trivial oscillatory function
\cite{VZ2}.

\end{document}